\documentclass[aps,prl,reprint,superscriptaddress]{revtex4-1} 
\usepackage{amsfonts}
\usepackage{amssymb}
\usepackage{amsthm,amsmath}
\usepackage{mathrsfs}
\usepackage{indentfirst}
\usepackage{multirow}
\usepackage{graphicx}
\usepackage{amsmath}
\usepackage[amssymb]{SIunits}
\usepackage{amssymb}
\usepackage{color}
\usepackage[usenames,dvipsnames]{xcolor}
\usepackage{natbib}
\usepackage[T1]{fontenc}
\usepackage[hidelinks, bookmarks=true]{hyperref}
\usepackage{bm}
\usepackage{booktabs}
\usepackage{setspace}

\begin{document}

\title{Sea ice aging by diffusion-driven desalination}

\author{Yihong Du}
\affiliation{New Cornerstone Science Laboratory, Center for Combustion Energy, Key Laboratory for Thermal Science and Power Engineering of MoE, Department of Energy and Power Engineering, Tsinghua University, China}

\author{Feng Wang}
\thanks{fengwang2023@tsinghua.edu.cn}
\affiliation{New Cornerstone Science Laboratory, Center for Combustion Energy, Key Laboratory for Thermal Science and Power Engineering of MoE, Department of Energy and Power Engineering, Tsinghua University, China}

\author{Enrico Calzavarini}
\affiliation{Univ.\ Lille, Unit\'e de M\'ecanique de Lille - J. Boussinesq (UML) ULR 7512, F-59000 Lille, France}

\author{Chao Sun}
\affiliation{New Cornerstone Science Laboratory, Center for Combustion Energy, Key Laboratory for Thermal Science and Power Engineering of MoE, Department of Energy and Power Engineering, Tsinghua University, China}
\affiliation{Department of Engineering Mechanics, School of Aerospace Engineering, Tsinghua University, China}

\date{\today}

\begin{abstract} 
Sea ice is a key component of the Earth's climate system, making its aging process an essential focus of current research.
The age of sea ice is closely linked to its thermal and mechanical properties, which govern its interactions with the surrounding environment. 
In this study, we combine experimental techniques and modeling to explore the full dynamical process of mushy ice growth and spontaneous aging in saline water, within a natural convective flow system.
We show that the aging of newly formed mushy ice in the present system, characterized by a gradual long-term reduction in porosity, is controlled by diffusion-driven desalination.
Moreover, we observe that the system eventually transits into a dense freshwater ice layer adjacent to a well-mixed saline water region.
The shape of the ice layer in this asymptotic state is well captured by numerical simulations of non-porous ice.
Our findings improve the understanding of the complex physics governing phase changes in aqueous systems and provide a framework for studying sea ice aging in laboratory settings, with implications spanning diverse natural and industrial applications.
\end{abstract}

\maketitle

In the cryosphere, sea ice evolves through distinct stages: first-year ice becomes second-year ice and eventually multi-year ice \cite{du2024physics}.
The stage of ice development (also called the ice age) directly influences its thermal and mechanical properties \cite{timco2010review}.
These properties play a critical role in how sea ice responds to environmental forces \cite{feltham2008sea, cenedese2023icebergs, roach2024physics}.
Changes in the spatial distribution of sea ice age \cite{maslanik2011distribution, korosov2018new} have important implications for the ice’s melting rate, areal extent and surface albedo, which can significantly affect a variety of climatic \cite{mori2019reconciled, screen2010central}, environmental \cite{peeken2018arctic} and ecological \cite{post2013ecological, arrigo2014sea} processes.

One distinct difference between first-year sea ice and older sea ice is their porosity.
Porosity is positively correlated with the salt content of the ice \cite{timco2010review}.
Previous studies have shown that sea ice undergoes significant desalination as it ages \cite{notz2009desalination, hunke2011multiphase, jardon2013full}.
Mushy layer theories have been proposed to describe the salt and heat transport as well as to estimate effective thermal properties \cite{feltham2006sea, notz2009desalination, worster2015sea, anderson2020convective}, which are applicable to mushy sea ice.
However, the dynamic aging process of mushy ice and its relationship with ice desalination still require further investigation, particularly through the integration of models and observations.

Laboratory studies have proven to be valuable for understanding the coupled physics of phase change and fluid flows in various scenarios \cite{du2024physics, esfahani2018basal, yang2024enhanced, wang2021growth, wang2021ice, wang2021equilibrium, weady2022anomalous, perissutti2024morphodynamics, yang2023bistability, wengrove2023melting, fang2024vibration}.
Many studies have explored the effects of brine convection in mushy ice on its structure and dynamics \cite{wettlaufer1997natural, worster1997convection, peppin2007steady, peppin2008steady, middleton2016visualizing, wells2019mushy, du2023sea}.
These studies focused primarily on the initial growth of ice and did not examine the long-term aging process.
More recently, laboratory experiments and simulations have revealed the rich dynamics of ice melting in salty water under various conditions
\cite{mondal2019ablation, wilson2023double, yang2023ice}.
However, in these studies, a pure ice state was adopted as an initial condition, and, as a consequence, the desalination of mushy ice was not involved.

In this work, we investigate the complete evolution of mushy ice in salty water within a vertical convection system using experiments and modeling.
Our findings reveal that ice desalination, driven primarily by salt diffusion and only to a minor extent by brine convection within the mushy ice, controls the aging process in the experimental system. 
This is also reflected by the decrease in the ice porosity.
To quantify this mechanism, we adopt an integral desalination model that captures well the experimental measurements.
Furthermore, we observe that the system eventually reaches a state similar to that of the solidification of a pure substance in the same configuration.
This work provides new measurements and insights to model brine drainage and sea ice aging.

\begin{figure*}[!htb]
\centering
\includegraphics[width=1\linewidth]{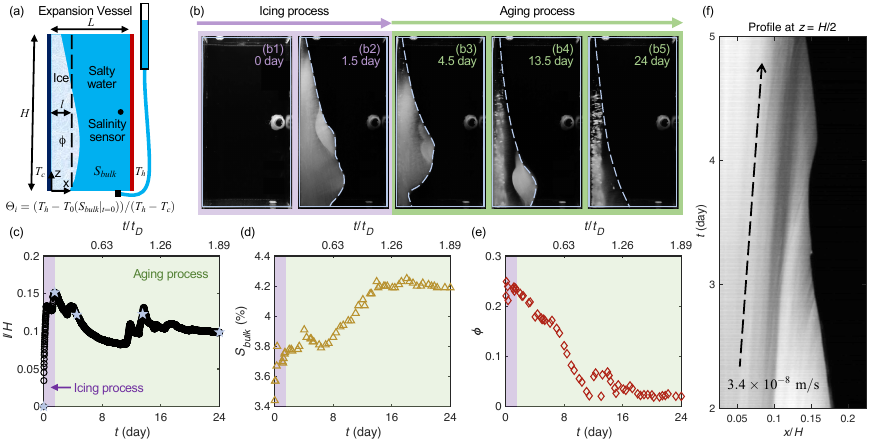}
\caption{Experiment on mushy ice evolution. 
(a) Sketch of the experimental setup.
In the experiment, an ice layer, characterized by its mean thickness $l$ and mean porosity $\phi$, forms and evolves in salty water within a vertical convection system.
The temperatures of the left cold plate and right hot plate are $T_c$ and $T_h$.
A salinity sensor measures the bulk salinity, $S_{bulk}$.
The initial superheat parameter is defined as $\Theta_i = (T_h - T_0(S_{bulk}|_{t=0})) / (T_h - T_c)$, where $T_0$ is the freezing point at the given salinity.
(b-f) Mushy ice evolution with $S_{bulk}|_{t=0}=3.44\%$, $T_c=-12.1$ $^{\circ}$C and $T_h=7.9$ $^{\circ}$C ($\Theta_i\approx 1/2$). 
The ice layer undergoes first a fast icing process (b1-b2, and purple regime in c-e), followed by a slow aging process (b3-b5 and green regime in c-e).
(b) Snapshots of the ice layer correspond to the blue stars in c.
(c-e) Temporal evolution of the spatially-averaged ice layer thickness $l$ made dimensionless by the system height $H$ (black circles in c), bulk salinity $S_{bulk}$ (yellow triangles in d) and the spatially-averaged ice layer porosity $\phi$ (red diamonds in e). 
The top axes show time nondimensionalized by the solutal diffusion time scale $t_D=l^{*2}/D \approx 1.1 \times 10^6$ s $\approx 12.7$ days, where $l^*$ is the mean ice thickness in the aging process and $D$ is the salt diffusivity in the brine.
(f) Space-time diagram for the ice layer profile along $z=0.5H$ from Day 2 to Day 5.
The internal structures of the mushy ice appears to migrate with a typical velocity of $3.0$ mm/day.
}
\label{fig1}
\end{figure*}

\emph{Experimental setup-} The experiments are performed in a quasi-two-dimensional vertical convection system \cite{ng2015vertical, shishkina2016momentum, wang2021ice, howland2023double} (Fig.\ref{fig1}a) with height $H=0.24$ m in $z$ direction, length $L=0.12$ m in $x$ direction, and width $W=0.06$ m.
The temperatures of left cold plate ($T_c$) and right hot plate ($T_h$) are maintained constant.
The system is initially filled with aqueous sodium chloride (NaCl) solution.
In the experiments, the bulk salinity $S_{bulk}$ is measured with a sensor and the ice front is photographed to determine the spatially averaged ice thickness in $x$ direction, $l$.
The spatially averaged ice porosity $\phi$ is calculated from conservation of salt, similar to \cite{wettlaufer1997natural}.
To quantify the effect of temperature, an initial superheat parameter is defined as $\Theta_i=(T_h-T_0(S_{bulk}|_{t=0}))/(T_h-T_c)$, where $T_0$ denotes the freezing point (see Supplementary Information \cite{SI} for details).

\begin{figure*}[!htb]
\centering
\includegraphics[width=0.9\linewidth]{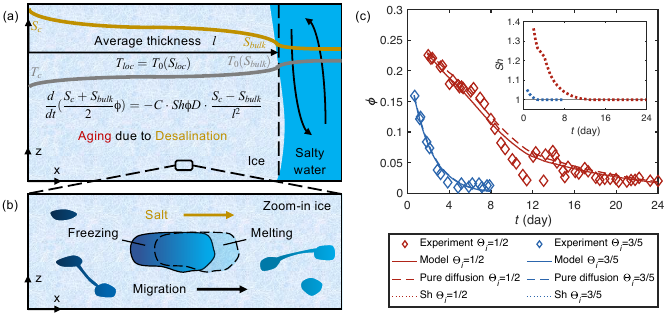}
\caption{Dynamics for mushy ice aging in salty water.
(a-b) Conceptual schematic of the desalination and aging of mushy ice for illustrative purposes.
(a) Sketch of the desalination of the whole ice layer.
The reduction of salt in the mushy ice equals salt removed through diffusion and advection.
The gray line sketches the variation of local temperature in $x$ direction, while the yellow line represents local salinity. 
The local temperature and salinity are related by the local thermal equilibrium in the mushy ice.
(b) Zoom-in sketch of the migration of pores inside the mushy ice. 
With the transport of salt, the colder (saltier) side freezes while the hotter (fresher) side melts, leading to the migration of ice internal structures.
(c) Integral desalination model (solid lines) and experiment (diamonds) results for the evolution of spatially-averaged ice layer porosity $\phi$, with initial superheat parameters $\Theta_i \approx 1/2$ (red) and $3/5$ (blue). 
The model predictions when advective transport is neglected by imposing $Sh\equiv1$ (dashed lines) are also attached for comparing the relative role of diffusion and advection. 
For $\Theta_i\approx3/5$, the blue dashed line overlaps largely with blue solid line.
The inset of panel c shows the corresponding time evolution of $Sh$ from the model (dotted lines).
}
\label{fig2}
\end{figure*}

\emph{Measurement results-} Fig.\ref{fig1}(b-f) display the ice layer evolution in a typical case with $S_{bulk}|_{t=0}=3.44\%$ and $\Theta_i \approx 1/2$ (see Movie.S1; see Movie.S2 and Supplementary Information \cite{SI} for $\Theta_i\approx3/5$).
The system first experiences a fast icing process at the left cold plate (purple regime in Fig.\ref{fig1}).
The ice layer grows in about 1.5 days to a maximal thickness (Fig.\ref{fig1}b1, b2, c).
As water freezes into ice, salt is not dissolvable in it.
Part of the salt is expelled into the salty water reservoir external to the ice layer, increasing the bulk salinity (Fig.\ref{fig1}d).
The other remains trapped in brine pockets in the ice matrix, leading to a highly porous ice layer (Fig.\ref{fig1}e) with a porosity roughly equivalent to that reported in previous studies \cite{petrich2006modelling, wettlaufer1997natural, worster1997natural, du2023sea}.
The rapid growth of mushy ice is driven by heat transfer across the ice and is controlled by the strength of the convection in bulk salty water \cite{wang2021growth}.
As the heat transfer on either side of the ice front balances \cite{du2023sea}, the rapid growth of the ice ceases.

Unlike in the freezing of fresh water or other pure liquids, the evolution of ice does not end here.
Without any change in the experimental conditions, the ice layer spontaneously enters an aging process (green regime in Fig.\ref{fig1}), which lasts until the system converges to the final equilibrium state (about 2 to 3 weeks in the present case).
Local advancements and retreats of the ice front occur (Fig.\ref{fig1}b3-b5), however, the average ice thickness varies with time only gradually and with an overall weak decreasing trend (Fig.\ref{fig1}c).
In contrast, the salinity of the bulk salty water increases (Fig.\ref{fig1}d).
The average ice porosity exhibits a slow yet noticeable decline (Fig.\ref{fig1}e) and the ice layer becomes gradually transparent, as is the case for pure ice (Fig.\ref{fig1}b3-b5).

Interestingly, the internal pore structures of the mushy ice appear to evolve and migrate towards the ice front over time (see Movie.S1).
Near the midheight of the ice layer, such migration is relatively obvious and essentially in $+x$ direction during Day 2 to Day 5 (see Movie.S3 for a zoom-in clip).
Fig.\ref{fig1}(f) shows how the photographed ice profile along $z=0.5H$ varies over time.
From this space-time diagram, a typical migration velocity can be identified to be about 3.0 mm/day.

\emph{Dynamics of ice aging-} The small variation in the average ice thickness suggests that the global thermal balance of the ice layer is preserved throughout the aging process.
Since it is not driven by thermal factors, what mechanism is responsible for the observed aging of mushy ice?

As heat diffusion is much faster than mass diffusion in salty water, mushy ice is in local thermal equilibrium \cite{feltham2006sea} during the aging process.
Figures \ref{fig2}(a,b) show the conceptual schematics of this process.
The local temperature always equals the local freezing/melting point.
The increase in temperature in $+x$ direction (gray line) leads to a decrease in local salinity (yellow line).
With such a salinity gradient, the salt is expelled from the mushy ice.
The salinity of the bulk salty water increases.
Meanwhile, excess water freezes inside the mushy ice, maintaining local salinity and resulting in the shrinkage of porosity.

As salt mass transfer controls the aging of ice, the characteristic timescale for this process is the solute diffusion time scale $t_D = l^{*2}/D$, where $l^*$ labels the spatio-temporally averaged ice thickness during the desalination process and $D$ is the salt diffusivity in the brine.
For the case with $\Theta_i\approx1/2$, $t_D$ is about $1.1\times10^6$ s $\approx12.7$ days.
As mentioned above, mass diffusion is much slower than heat diffusion.
In fact, the Lewis number $Le=\kappa/D$ is roughly 200, where $\kappa$ is the brine thermal diffusivity.
This explains the significant difference between the time scales of the icing process and the aging process (see Fig.\ref{fig1}c-e).

The apparent velocity scale for salt diffusion can be estimated by dividing the diffusive salt flux $D(S_c-S_{bulk})/l$ with the reference salinity $(S_c+S_{bulk})/2$, i.e., $v_D=2D(S_c-S_{bulk})/(l(S_c+S_{bulk}))$ \cite{notz2009desalination, rempel2001possible}, where $S_c$ is the salinity at the cold plate determined by its temperature $T_c$, and the bulk salinity $S_{bulk}$ is approximately the salinity at the ice front, assuming sufficiently strong convective mixing in the bulk.
Considering the pore structures inside the mushy ice (Fig.\ref{fig2}b), the saltier (colder) side freezes while the fresher (hotter) side melts with the diffusion of salt.
Therefore, the apparent velocity scale for salt diffusion also reflects the typical migration velocity of internal pore structures \cite{hoekstra1965migration}.
For the case in Fig.\ref{fig1}(f), it is about $1.8$ to $2.3$ mm/day, close to the observed migration velocity.
The remaining difference might come from the influence of non-uniform ice thickness and porosity on the global and local desalination characteristics, as well as the influence of brine convection.

Noteworthy, the decrease in the global salt content in the mushy layer is directly related to the salt released into the bulk salty water as the pores migrate to the ice front, as the result of an effective diffusive process controlled by the evolving salinity gap across the also evolving ice layer (see Fig.\ref{fig2}a).
The temporal evolution of this process can be captured by the following integral desalination model:
\begin{equation}
    \begin{aligned}
        \frac{d}{dt}\left( \frac{S_c+S_{bulk}}{2} \phi\right)&=- C\cdot Sh v_D \cdot \phi\frac{S_c+S_{bulk}}{2l} \\&=- C\cdot  Sh\phi D\cdot \frac{S_c-S_{bulk}}{l^2},
    \end{aligned}
    \label{eq1}
\end{equation}
where $S_c$ is the salinity at the cold plate determined by $T_c$ in the experiment; $S_{bulk}$ is the bulk salty water salinity and $l$ is the average ice thickness, both extracted from experiments; $D$ is the salt diffusivity; $Sh$ is the Sherwood number representing the dimensionless mass transfer efficiency due to convection in mushy ice, whose estimation is shown in End Matter; parameter $C$ compensates for influences of non-uniform ice thickness and porosity (see details in Supplementary Information \cite{SI}).
With a given initial value of porosity $\phi$, the temporal evolution of $\phi$ can be explicitly solved by the model.

We choose $C=1.1$ for $\Theta_i\approx1/2$ (red) and $C=2.2$ for $\Theta_i\approx3/5$ (blue) to best fit the experimental data.
Fig.\ref{fig2}(c) compares the model (lines) and experimental (diamonds) results for the average porosity $\phi$ over time. 
The model agrees well with the experimental results, demonstrating that it effectively captures the desalination mechanism that governs the aging dynamics of the ice.
Convective salt transport inside the ice occurs only at the beginning of the process and plays a much more limited role in the desalination and ice aging within the parameter regime of the current system. 
This is evident from the inset of Fig.\ref{fig2}(c) and also from the small differences between the solid lines (including advective transport) and the dashed lines (assuming pure diffusion) in Fig.\ref{fig2}(c) main panel, which is even negligible when $\Theta_i\approx3/5$ (blue).
This result explains why the duration of the aging process and the migration velocity of the observed patterns are comparable to the timescale and the velocity scale of salt diffusion, as shown in Fig.\ref{fig1}(c-e) and discussed above. 
A weakening convection in mushy ice when $\Theta_i\approx1/2$ also explains the overall decrease in ice thickness through the aging process, whereas the negligible convection in mushy ice explains the very small changes in ice thickness for $\Theta_i\approx3/5$ \cite{du2023sea}.
 
\begin{figure}[!tb]
\centering
\includegraphics[width=0.9\linewidth]{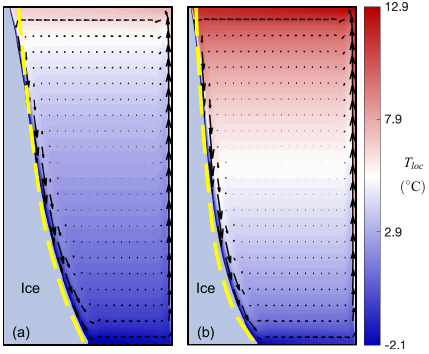}
\caption{Asymptotic equilibrium state of the system. 
The final ice front morphologies from the experiment (yellow dashed line) are compared with those from direct numerical simulation of a positively buoyant pure liquid freezing under the same conditions (thin black line). 
The color shows the temperature field in the liquid phase, while the small black arrows indicate the velocity field, both from the simulation.
The ``ad hoc" positively buoyant liquid possess the same thermal properties as salty water with the bulk salinity measured at the end of the experiments.
(a) $\Theta_i\approx 1/2$. 
(b) $\Theta_i\approx 3/5$.
}
\label{fig3}
\end{figure}

Naturally, the local formation and melting of more porous or less porous regions within the ice layer influence the average porosity.
A detailed explanation of the ice morphology evolution is highly challenging as it requires detailed and nonintrusive measurements of the velocity, temperature and salinity fields in the mushy ice and in the bulk salty water.
Still, it is worth pointing out that the final form of the ice front (yellow dashed line in Fig.\ref{fig3}) overlaps remarkably well with the one obtained from direct numerical simulations of the solidification of a pure liquid (thin black line in Fig.\ref{fig3}). 
In the simulations, we use the same physical properties (same salinity and same freezing point) for the fluid as those of the salty water in the final state of the experiment.
The results suggest that the final ice layer in the experiment is, in practice, a pure ice layer, and the bulk salty water eventually becomes well mixed with uniform salinity.
Vertical convective flow dominates in the final bulk salty water, with temperature imbalances being the sole contributor to density differences and buoyancy.
We refer to Supplementary Information \cite{SI} for more details.

Overall, the combined experimental, theoretical and numerical results provide a consistent understanding of the dynamics of ice aging in salty water. 
They highlight the critical role of thermal gradients in salinity and ice porosity evolution, and the eventual transformation of mushy ice into solid ice.

\emph{Conclusion and outlook-} In summary, we study the complete evolution of ice in salty water in a vertical convective flow system, with a focus on the dynamics of its spontaneous aging. 
We show that in the present system and parameter regime, the observed aging of mushy ice is driven primarily by diffusive desalination and to a minor extent by convective desalination. 
The long-term decline of ice porosity can be well predicted using an integral model that quantifies the salt flux out of the ice layer. 
In addition, we find that the final state of the system, consisting of a dense ice layer and a well-mixed salty water zone, is highly consistent with the freezing of a positively buoyant pure liquid under the same conditions. 

Our results advance the understanding of the coupled dynamics of fluid flow, mass and heat transfer, and phase change in aqueous solution systems.
In natural scenarios of sea ice, rapid freezing and melting are dominating in case of drastic thermal disequilibrium, while diffusion-driven desalination and aging of sea ice are slow and occur in the long term.
This indicates the need to adopt a multi-layer model for multi-year sea ice as it may experience multiple freezing and melting during its aging.
Noteworthy, there can also exist stronger desalination processes such as flushing or brine convection, accelerating the aging of sea ice.
Although the aging of real-world sea ice is more complex, our findings provide a valuable example of how this process can be investigated and parameterized through laboratory studies. 
Future work could explore different flow geometries, and incorporate the effects of multicomponent salts and impurities typically found in natural seawater. 
It would also be important to investigate how the aging process affects the overall heat transfer in ice and bulk salty water.
Combining field measurements with laboratory studies will provide new insights into the natural aging of sea ice, refine parameterizations and enable more accurate integration into climate models, helping bridge the gap between controlled experiments and the complexities of sea ice dynamics in natural environments.\\

\noindent
{\bf Acknowledgements:} We thank Rui Yang for insightful discussions.
This work is supported by NSFC Excellence Research Group Program for ‘Multiscale Problems in Nonlinear Mechanics’ (No. 12588201), the Natural Science Foundation of China under Grant No. 12402321, the New Cornerstone Science Foundation through the New Cornerstone Investigator Program and the XPLORER PRIZE, and Shuimu Tsinghua Scholar Program under Grant No. 2023SM038.\\

%\bibliographystyle{apsrev4-2}
%\bibliography{ref.bib}


\bigskip
\onecolumngrid

\section*{End Matter}

\twocolumngrid
\noindent
{\bf Derivation of the integral model}

Neglecting salt dissolution in ice and the density difference between brine and ice, the governing equation for the local salt conservation in the porous ice reads \cite{feltham2006sea, notz2009desalination},
\begin{equation}
\begin{aligned}
    \partial (S_{loc}\phi_{loc})/{\partial t}=\nabla\cdot(\phi_{loc}D\nabla S_{loc})-\phi_{loc}\textbf{u}\cdot \nabla S_{loc},
\end{aligned}   
\label{eq2}
\end{equation}
where $S_{loc}$ is the local salinity in the brine, $\phi_{loc}$ is the local porosity, $D$ is the salt diffusivity and $\mathbf{u}$ is the brine velocity.
While solving Eq.\ref{eq2} directly is complex, its physical interpretation is straightforward: The rate of change for the mass of salt is determined by the diffusive and advective transport of salt. 

We estimate the rate of change for the total mass of salt in ice as,
\begin{equation}
    \begin{aligned}
        \frac{d}{dt}\left(\frac{S_c + S_{bulk}}{2} \phi\right)\cdot \rho_mlHW,
    \end{aligned}
    \label{eq3}
\end{equation}
where $(S_c + S_{bulk}/2)$ is approximately the average salinity of the brine in the mushy ice, $\phi\cdot \rho_mlHW$ is the mass of brine in the mushy ice.
Here we neglect the time dependence of $\rho_m$ and $l$, since their variations are small.

Assuming that $S_{bulk}$ is roughly the salinity at the ice front, the diffusive and advective transport of salt from ice into bulk salty water are estimated jointly as,
\begin{equation}
    \begin{aligned}
        C\cdot Sh \phi D \cdot\frac{S_c - S_{bulk}}{l}\cdot \rho_mHW,
    \end{aligned}
    \label{eq4}
\end{equation}
where parameter $C$ is introduced to compensate for the non-uniform ice thickness and porosity.
The Sherwood number $Sh=Jl/(D(S_c-S_{bulk}))$ quantifies the dimensionless mass transport efficiency due to convection (sum of diffusion and advection), where $J$ is the salt flux.

Finally, by equating Eq.\ref{eq3} and Eq.\ref{eq4}, we obtain the integral model (Eq.\ref{eq1}).

\begin{table}[b]
    \centering
    \begin{tabular}{cl}
    \toprule
    Symbol& Physical meaning\\
    \midrule
    $C$& Correcting coefficient\\
    $D$& Diffusivity of salt in brine\\
    $g$& Gravitational acceleration\\
    $H$& Height of experiment system\\
    $J$& Salt flux\\
    $K$& Permeability\\
    $L$& Length of experiment system\\
    $l$& Spatially-averaged ice layer thickness\\
    $l^*$& Mean ice thickness in aging process\\
    $Le$& Lewis number\\
    $Ra$& Rayleigh number in mushy ice\\
    $S_{bulk}$& Salinity of bulk salty water\\
    $S_c$& Salinity at the cold plate\\
    $S_{loc}$& Local salinity\\
    $Sh$& Sherwood number\\
    $T_c$& Left cold plate temperature\\
    $T_h$& Right hot plate temperature\\
    $T_0(S)$& Freezing/melting point at salinity $S$\\
    $t$& Time\\
    $t_D$& Solutal diffusion time scale\\
    $\textbf{u}$& Local velocity in brine in mushy ice\\
    $v_D$& Apparent velocity scale of salt diffusion\\
    $W$& Width of experiment system\\
    $\kappa$& Brine thermal diffusivity \\
    $\nu$& Brine kinetic viscosity\\
    $\phi$& Spatially-averaged ice layer porosity\\
    $\phi_{loc}$& Local ice porosity\\
    $\rho_0$& Salty water density at ice front\\
    $\rho_c$& Brine density at the cold plate\\
    $\rho_m$& Brine density in the mushy ice\\
    $\Theta_i$& Initial hot plate superheat parameter\\
    \bottomrule
    \end{tabular}
    \caption{Summary of symbols and their physical meanings.}
    \label{table1}
\end{table}

\noindent
{\bf Estimation of the Sherwood number}

The convection in mushy ice in our study is mainly driven by salinity gradient.
The scaling of mass transfer is comparable to that of heat transfer for convection driven by temperature gradient \cite{bejan1985heat}.
We obtain the estimation for $Sh$ by fitting the experimental data in \cite{schneider1963investigation}:  
\begin{equation}
    \begin{aligned}
    Sh=1+0.04\times(l/H)^2LeRa.
    \end{aligned}
\end{equation}

$Le$ is the Lewis number, representing the ratio of thermal diffusivity $\kappa$ to the mass diffusivity of salt $D$,
\begin{equation}
    \begin{aligned}
        Le=\kappa/D.
    \end{aligned}
\end{equation}

$Ra$ is the Rayleigh number, which parametrizes the relative role of the buoyancy intensity driving the convection with respect to the thermal and kinetic dissipation,
\begin{equation}
    \begin{aligned}
        Ra=\frac{KgH}{\kappa \nu}\cdot \frac{\rho_c-\rho_0}{\rho_0},
    \end{aligned}
\end{equation}
where $g$ is gravitational acceleration, $H$ is the system height, $\nu$ is the kinetic viscosity, $\rho_c$ is the brine density at cold plate and $\rho_0$ is the brine density at the ice front.

The permeability $K$ (unit: m$^2$) in the definition of $Ra$ depends on the structure of the ice. It is estimated as a function of the average ice porosity $\phi$, modified from the empirical equation in previous study \citep{petrich2006modelling},
\begin{equation}
	K=\left\{
	\begin{aligned}
		&0 ,& \phi < 0.054, \\
		&2\times10^{-11}(\phi-0.054)^{1.2} ,& \phi > 0.054.
	\end{aligned}
	\right .
\end{equation}

This estimate is in good agreement with previous laboratory and field results \citep{ petrich2006modelling}.

\noindent
{\bf Table of symbols}
\noindent \\
A summary of the symbols and their physical meanings are provided in Table.\ref{table1}, for the convenience of reading.
%\end{document}

%\begin{document}
\setcounter{figure}{0} 
\setcounter{equation}{0} 
\renewcommand\thefigure{S\arabic{figure}}
\renewcommand\theequation{S\arabic{equation}}
\onecolumngrid

\section*{Sea ice aging by diffusion-driven desalination: Supplementary information}

\subsection{1. Experimental setup and data measurements}

The experiments are performed in a quasi-two-dimensional vertical convection (VC) system (see Fig.1 in the main text). 
The effective experimental volume is of height $H=0.24$ m in $z$ direction, length $L=0.12$ m in $x$ direction and width $W=0.06$ m.
The VC system consists of two parallel vertical copper plates and polymethyl methacrylate (PMMA) sidewalls. 
An expansion vessel is connected to the system to compensate and monitor the volume change resulting from thermal expansion and phase change.
In the experiments, the left cold plate is of temperature $T_c$ and the right hot plate is of temperature $T_h$ with $T_c<T_h$. 
Their temperatures are measured with the thermistors embedded in them and maintained constant with circulating baths. 
The system is placed in an insulated box. 
The temperature inside the box is controlled at $0.5(T_c+T_h)$ to minimize the influence of heat transfer through the side walls.
The working fluid is aqueous NaCl (sodium chloride) solution of initial salinity $S_{bulk}|_{t=0}$.
Before the preparation of the solution, water is boiled twice to be degassed.

As the experiment begins, a porous mushy ice layer forms on the left cold plate, which grows to the right and evolves until the final state. 
Meanwhile, the horizontal temperature difference leads to nonuniformity of the salty water density, resulting in buoyancy-driven vertical convection in the system.

In this work, we focus mainly on two cases: (1) $S_{bulk}|_{t=0}=3.44\%$, $T_c=-12.1$ $^{\circ}$C, $T_h=7.9$ $^{\circ}$C and (2) $S_{bulk}|_{t=0}=3.25\%$, $T_c=-12.1$ $^{\circ}$C, $T_h=12.9$ $^{\circ}$C.
In both cases, $S_{bulk}|_{t=0}$ is chosen to be around the average ocean salinity, which is about $3.5\%$.
To control the initial supercooling of the left cold plate, $T_c$ is chosen as $-12.1$ $^{\circ}$C in both cases, approximately $10$ $^{\circ}$C below the freezing point of salty water at the initial salinity $T_0(S_{bulk}|_{t=0})$.
To examine the effect of temperature, we vary $T_h$ to examine the effect of temperature.
An initial superheat parameter is defined as $\Theta_i=(T_h-T_0(S_{bulk}|_{t=0}))/(T_h-T_c)$, which is $\Theta_i\approx1/2$ in Case (1) and $\Theta_i\approx3/5$ in Case (2).

We measure three properties in the experiments: the spatially averaged ice layer thickness $l$, the bulk salinity in the salty water $S_{bulk}$, and the spatially averaged ice layer porosity $\phi$.
The spatially averaged ice layer thickness $l$ is determined by identifying the ice front (the boundary between the ice layer and the bulk salty water) in the photos taken during the experiments.
The bulk salinity in salty water $S_{bulk}$ is measured using a salinity sensor placed at the midheight of the experimental cell near the hot plate.

The spatially-averaged ice layer porosity $\phi$ is calculated by monitoring the volume changes of the salty water in the expansion vessel and solving the conservation of salt in the system,
\begin{equation}
    \begin{aligned}
(S_{bulk}\rho_{bulk})|_{t=0}=\frac{S_c+S_{bulk}}{2}\rho_{brine}\cdot \phi\frac{l}{L}+S_{bulk}\rho_{bulk}\cdot (1-\frac{l}{L}+\frac{\Delta v_e}{LHW}).
    \end{aligned}
    \label{eqs1}
\end{equation}
Here, $S_{bulk}$ is the bulk salinity, $\rho_{bulk}$ is the density of bulk salty water, $(S_c+S_{bulk})/2$ is the average salinity of brine in the mushy ice where $S_c$ is the salinity at the cold plate determined by local thermal equilibrium, $\rho_{brine}$ is the average density of brine in the mushy ice, $\phi$ is the spatially averaged ice porosity, $l$ is the spatially averaged ice layer thickness, $L$, $H$ and $W$ are the length, height and width of the experimental cell, and $\Delta v_e$ is the change in the salty water volume in the expansion vessel relative to the initial volume.
This method of measuring average ice porosity by solving the conservation of salt has been validated in previous studies \cite{wettlaufer1997natural1, du2023sea1}.
The ice porosity is measured to be about 0.2 to 0.25 during the icing process, which is equivalent to that reported in many studies \cite{petrich2006modelling1, wettlaufer1997natural1, worster1997natural1, du2023sea1}.
This proves the reliability of the calculation.

\subsection{2. Experiment results for $\Theta_i\approx3/5$}

\begin{figure}[!b]
\centering
\includegraphics[width=0.8\linewidth]{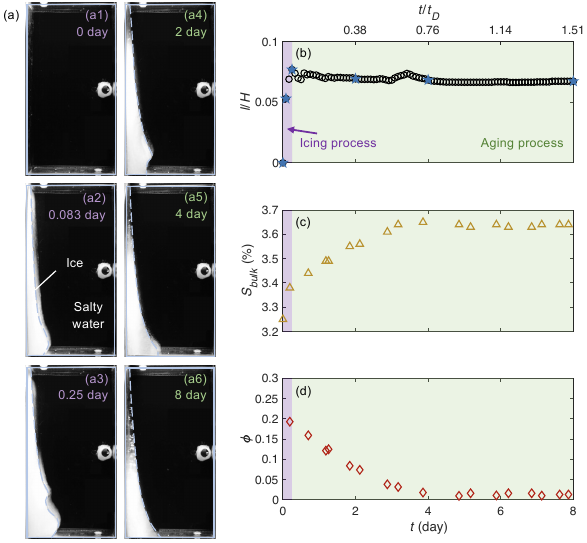}
\caption{Mushy ice evolution in an experiment with $S_{bulk}|_{t=0}=3.25\%$, $T_c=-12.1$ $^{\circ}$C, and $T_h=12.9$ $^{\circ}$C ($\Theta_i\approx 3/5$).
Compared to the case with $\Theta_i\approx 1/2$ shown in Fig.1 of the main text, the ice layer underwent a similarly fast icing process (a1-a3 and the purple regime in b-d), followed by a slow aging process (a4-a6 and the green regime in b-d). 
However, fluctuations in the ice front were fewer, and the average ice thickness was more stable.
(a) Snapshots of the ice layer correspond to the blue stars in b.
(b-d) Temporal evolution of the spatially averaged ice layer thickness $l$ nondimensionalized by the system height $H$ (black circles in b), bulk salinity $S_{bulk}$ (yellow triangles in c), and spatially averaged ice layer porosity $\phi$ (red diamonds in d). 
The top axis in b show time nondimensionalized by the solutal diffusion time scale $t_D=l^{*2}/D \approx 4.6 \times 10^5$ s $\approx 5.3$ days, where $l^*$ is the mean ice thickness in the aging process and $D$ is the salt diffusivity in the brine.
}
\label{figs1}
\end{figure}

Fig.\ref{figs1} shows the ice evolution in an experiment with $S_{bulk}|_{t=0}=3.25\%$, $T_c=-12.1$ $^{\circ}$C, and $T_h=12.9$ $^{\circ}$C ($\Theta_i\approx 3/5$), see Movie.S2 in the Supplementary Materials for the video. 
The ice exhibits similar dynamics to those in the case of $\Theta_i\approx 1/2$ (Fig. 1 in the main text).
The ice layer first undergoes a fast icing process (a1-a3 and the purple regime in b-d). During the icing process, the ice layer thickens rapidly, the bulk salinity increases, and the ice porosity is high.
Then, the ice layer enters the aging process, characterized by a slow decrease in ice porosity. During the aging process, the ice layer thickness changes little, and the bulk salinity increases. At the end of the aging process, the ice porosity becomes very low ($\approx 0$).
The morphology of the ice front is also similar to that of the case with $\Theta_i\approx 1/2$ (Fig.1 in the main text). 
A flat and inclined ice front occupies the upper part of the ice layer throughout the experiment. 
The lower part of the ice layer is thinner near the bottom during the icing process, and local advancements and retreats of the ice front occur during the aging process. 
Finally, a flat and inclined ice front dominates the morphology of the entire ice layer at the end of the aging process.

Compared to the case with $\Theta_i\approx 1/2$, the ice layer in the case with $\Theta_i\approx 3/5$ is overall thinner. 
Fluctuations in the ice front are fewer and the average ice thickness is more stable. 
Meanwhile, the aging process lasts for a shorter duration, yet remains comparable to the solutal diffusion time scale $t_D$.

\subsection{3. Choice of $C$ in the model}

In the 1D integrated model, the salinity gradient within the ice layer is defined as the salinity difference across the ice layer $(S_c-S_{bulk})$ over the average ice thickness $l$.
The effective salt diffusivity within the porous ice is defined as the product of the average ice porosity $\phi$ and the salt diffusivity $D$.
Both the non-uniform ice thickness and porosity may influence the local characteristics of the salt transport. 
Therefore, according to the observations in experiments (as shown in Fig.1(b) in the main text), we need to introduce an empirical parameter $C$ in the 1D integrated model to compensate the effect of non-uniform ice thickness and porosity.

\begin{figure*}[!htb]
\centering
\includegraphics[width=0.8\linewidth]{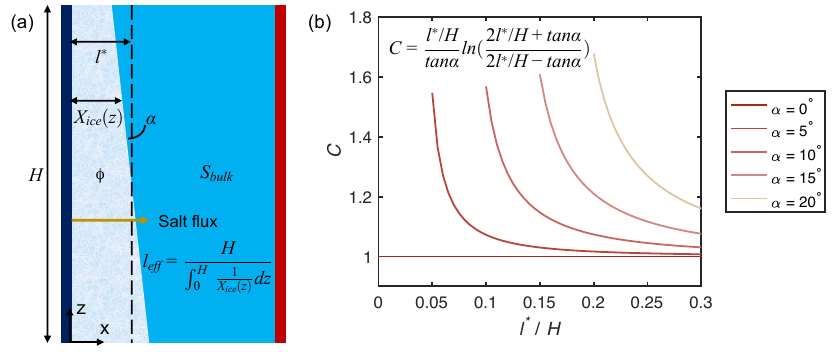}
\caption{Effect of non-uniform ice thickness on the desalination and aging process.
(a) Sketch of a simplified scenario illustrating the effects of non-uniform thickness. 
The salinity gradient and the induced salt flux at different vertical heights is assumed to depend on the local ice thickness $X_{ice}(z)$, leading to an effective thickness for salt diffusion $l_{eff}$.
The ice layer is assumed to be non-deformable, of thickness $l^*$ (mean ice thickness in the aging process) and of a flat ice front inclined at angle $\alpha$ to the direction of gravity.
(b) Compensation parameter $C$ as function of $l^*/H$ at different $\alpha$. In this case, $C$ accounts for the enhancement of desalination due to non-uniform ice thickness.
}
\label{figs2}
\end{figure*}

As revealed by the previous work \cite{wang2021ice1}, in the vertical convection system, the ice tends to be shaped into a wedge with a slope.
Here, we examine a simplified scenario to illustrate the influences of non-uniform thickness, see Fig.\ref{figs2}.
We consider the salinity gradient in ice and the induced salt flux at different vertical heights to depend on the local ice thickness $X_{ice}(z)$ rather than on the mean ice thickness $l$.
The effective diffusivity is still estimated as the product of the average ice porosity $\phi$ and the salt diffusivity $D$.
This leads to an effective thickness for salt diffusion, the harmonic mean value of ice thickness,
\begin{equation}
\begin{aligned}
    l_{eff}=\frac{H}{\int_0^H(1/X_{ice}(z))dz},
\end{aligned}
\label{eqs2}
\end{equation}
where $H$ is the system height.
In this case, the amplification of desalination due to desalination is $l/l_{eff}$.
Further,  we neglect the ice deformation during the aging process, assuming an ice layer of thickness $l^*$ (mean ice thickness in the aging process) and of a flat ice front inclined at angle $\alpha$ to the direction of gravity, see Fig.\ref{figs2}(a),
\begin{equation}
\begin{aligned}
    X_{ice}(z)=-\tan\alpha\cdot z+l^*+\frac{\tan\alpha\cdot H}{2}.
\end{aligned}
\label{eqs3}
\end{equation}
Combining Eq.\ref{eqs2} and Eq.\ref{eqs3}, we obtain,
\begin{equation}
\begin{aligned}
    C=\frac{l^*/H}{\tan\alpha}\ln(\frac{2l^*/H+\tan\alpha}{2l^*/H-\tan\alpha}).
\end{aligned}
\label{eqs4}
\end{equation}

Fig.\ref{figs2}(b) shows $C$ as function of $l^*/H$ at different $\alpha$.
It turns out that $C$ increases with $\alpha$ and decreases with $l^*/H$.
For the experiment with $\Theta_i=1/2$, $\alpha\approx10^\circ$ and $l^*/H\approx0.11$, corresponding to $C\approx1.4$.
For the experiment with $\Theta_i=3/5$, $\alpha\approx7^\circ$ and $l^*/H\approx0.07$, corresponding to $C\approx1.6$.
The results in Fig.\ref{figs2} clearly show how the non-uniform ice thickness may affect the desalination efficiency.
Furthermore, it should be noted that the derivation of Eq.\ref{eqs4} here is based on many assumptions and simplifications.
For example, the ice thickness varies with time in the experiments. 
The ice front is curved and also varies over time.
Therefore, it is quite challenging to calculate the explicit value of the parameter $C$ based on the morphodynamics of the ice layer.

At the same time, in the experiments, the ice porosity can also be non-uniform, as indicated by the non-uniform transparency of the ice layer in Fig.1(b) in the main text, which may be another major factor that influences the desalination process.
For example, as there is no salt flux through the cold plate, ice at the colder side may preferentially lose its salts and become less porous.
This makes the ice porosity lower near the cold plate and higher near the ice front.
When convection occurs in the mushy ice, it may bring fresher water to the upper region of the ice and enhance freezing in the pore structures.
The local porosity is also expected to be lower where the ice is thinner, as the salinity gradient is larger and desalination is stronger.
These may lead to porosity variations in the vertical direction.
The detailed porosity distribution in ice can be very complex, making it difficult to have a feasible theoretical model or numerical results for the effective porosity and for $C$.

\begin{figure*}[!htb]
\centering
\includegraphics[width=0.8\linewidth]{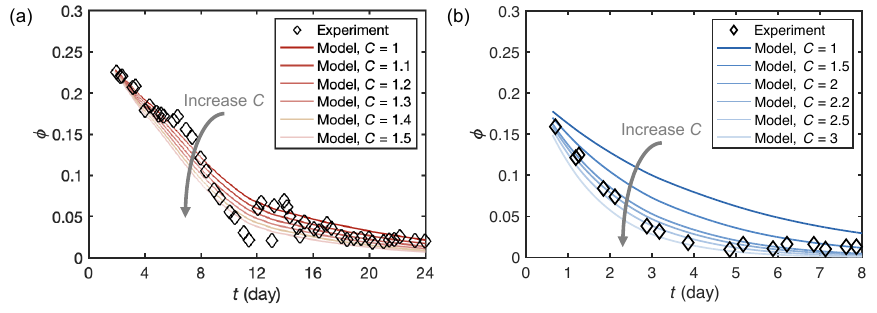}
\caption{Desalination model (lines) and experiment (black diamonds) results for ice porosity during the aging of mushy ice with different correction parameter $C$. (a) $\Theta_i=1/2$ and (b) $\Theta_i=3/5$.
}
\label{figs3}
\end{figure*}

As non-uniform ice thickness and non-uniform ice porosity can have complicated effects on desalination and aging of mushy ice that is difficult to parameterize, we adopt the empirical parameter $C$ to account for these effects and choose its values to best fit the experiment results in the manuscript.
The chosen values are $C=1.1$ for $\Theta_i=1/2$ and $C=2.2$ for $\Theta_i=3/5$.
Qualitatively, they are consistent with the calculation results in Fig.\ref{figs2}.
Below we attach the model result with different values of $C$ in Fig.\ref{figs3} for reference.
It should be noticed that, although there is an empirical parameter $C$, the model is able to capture the main mechanism of mushy ice aging in our experiments: The decrease in ice porosity is controlled by diffusion-driven desalination.

\subsection{4. Numerical simulation}

We perform a two-dimensional numerical simulation on the freezing of a hypothetical positively buoyant liquid using the \textsc{ch4-project} code \cite{calzavarini2019eulerian1}.
The code adopts the Lattice-Boltzmann method (LBM) to capture the rich dynamics of vertical natural convection and heat transport (see \cite{jiang2019robustness1} for a previous validation against experiments), and adopts the enthalpy method to model the physics of phase change at the solid-liquid interface.

The physical height and length of the computational domain are the same as those of the experimental system, i.e., $H \times L = 0.24$ m $\times$ $0.12$ m, with a resolution of $2048 \times 1024$.
The left and right boundary conditions are non-slip and isothermal, with the left plate temperature set to $T_c = -12.1$ $^{\circ}$C and the right plate temperatures to $T_h = 7.9$ and $12.9$ $^{\circ}$C, corresponding to the experiments.
The top and bottom boundary conditions are non-slip and adiabatic.
The solid-liquid boundary conditions are non-slip and follow the Stefan condition \cite{gupta2017classical1}.

To compare the results between experiment and simulation, we assign to this hypothetical liquid the same thermal properties as an aqueous NaCl solution of the final bulk salinity in the experiments, including the equation of state \cite{gebhart1977new1} and freezing/melting point \cite{hall1988freezing1}:
\begin{equation}
    \begin{aligned}
        \rho_{loc}=999.972 (1+0.8046S) [1-9.297\times 10^{-6} (1-2.839S) |T_{loc}-3.98 (1-52.66S)|^{1.895}],
    \end{aligned}
    \label{eqs5}
\end{equation}
\begin{equation}
    \begin{aligned}
       T_0=-60.37S-581.23S^3,
    \end{aligned}
    \label{eqs6}
\end{equation}
where $\rho_{loc}$ is the local salty water density (unit: kg/m$^3$), $S$ is the salinity ($S=4.2\%$ for $\Theta_i\approx1/2$ and $S=3.6\%$ for $\Theta_i\approx3/5$), $T_{loc}$ is the local temperature (unit: $^\circ$C), $T_0$ is the freezing and melting point (unit: $^\circ$C), and $|...|$ denotes the absolute value.

We recognize that equilibrium is reached when the average solid layer thickness varied by less than 2\% during the last $5 \times 10^7$ time steps (corresponding to $7.2\times10^3$ s $\approx 0.08$ day for $\Theta_i\approx1/2$ and $7.7\times10^3$ s $\approx 0.09$ day for $\Theta_i\approx3/5$), with the entire simulations lasting for $3 \times 10^8$ time steps (corresponding to $4.3\times10^4$ s $\approx 0.5$ day for $\Theta_i\approx1/2$ and $4.6\times10^4$ s $\approx 0.54$ day for $\Theta_i\approx3/5$).
For more detailed information on the governing equations for mass conservation, fluid flow, heat transfer, and phase change, we refer to \cite{esfahani2018basal1, wang2021growth1, wang2021ice1}, where the code has been extensively validated against experiments and theoretical models.

\begin{figure*}[!htb]
\centering
\includegraphics[width=1\linewidth]{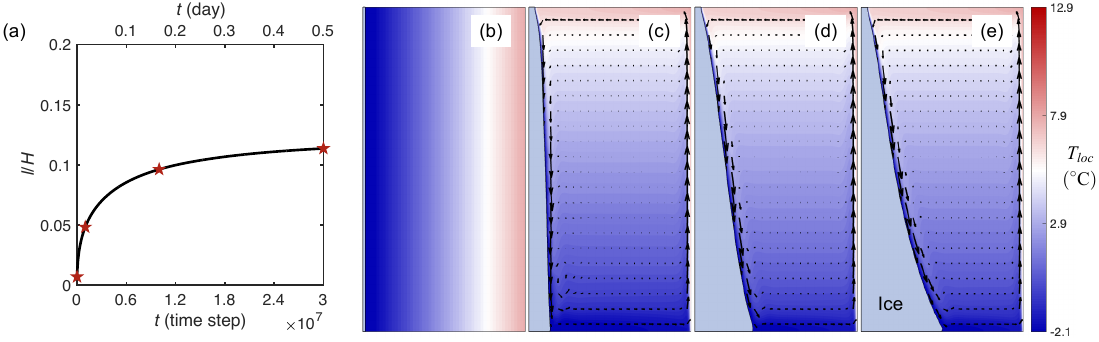}
\caption{Freezing dynamics of a positively buoyant pure liquid from direct numerical simulation.
The environmental conditions are the same as those in the experiment with $\Theta_i=1/2$.
The hypothetical liquid is assigned the same thermal properties as salty water of the final bulk salinity in the experiment at $\Theta_i=1/2$.
The entire simulation lasts for $3 \times 10^8$ time steps, corresponding to $4.3\times10^4$ s $\approx 0.5$ day.
(a) Temporal evolution of the dimensionless ice layer thickness $l/H$.
(b-e) Ice layer, temperature field (color) and velocity field (arrows) in the liquid when (b) $t=0$, (c) $t=1\times10^6$ time steps $\approx 0.017$ day, (d)  $t=1\times10^7$ time steps $\approx 0.17$ day, (e)  $t=3\times10^7$ time steps $\approx 0.5$ day, corresponding to the red stars in panel a.
}
\label{figs4}
\end{figure*}

Fig.\ref{figs4} shows the growth of ice layer in the simulation for $\Theta_i=1/2$.
The simulation starts with a thin layer of ice (Fig.\ref{figs4}a,b).
The initial temperature in ice and in liquid is set to increase linearly (Fig.\ref{figs4}b).
As the simulation starts, the temperature and velocity fields quickly converge to that of vertical convection \cite{shishkina2016momentum1, wang2021ice1}, as shown in Fig.\ref{figs4}(c-e).
Unlike the case in salty water, the average ice layer thickness increases monotonically and approaches the equilibrium thickness quickly (Fig.\ref{figs4}a).
The ice front morphology remains similar during this process (Fig.\ref{figs4}c-e).
While experiencing very different evolutions, the final shapes of the ice front in the experiment and in the simulation overlap remarkably well, see Fig.3 in the main text.
This suggests that the final ice layer in the experiment is in practice a pure ice layer.
The bulk salty water eventually becomes well mixed and dominated by vertical convection.
To save on costs, the simulation for $\Theta_i=3/5$ inputs the final state for $\Theta_i=1/2$ as initial condition, with the thermal properties modified.
Meanwhile, the maximal velocity in the final state of the simulations is $3.4$ mm/s for $\Theta_i=1/2$ and $4.2$ mm/s for $\Theta_i=3/5$.
The Reynolds numbers are in the $10^2$ order of magnitude range.
These velocities and Reynolds numbers are also consistent with those in previous work on vertical convection \cite{shishkina2016momentum1}.

\subsection{5. Flow chart on the measurement and modeling of porosity}

\begin{figure}[!htb]
\centering
\includegraphics[width=0.8\linewidth]{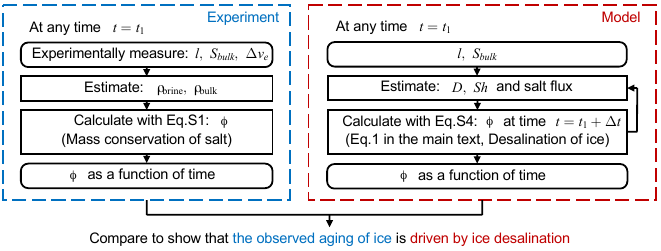}
\caption{Flow chart on how the average ice porosity $\phi$ as a function of time is measured in the experiment and estimated with the model.
In the experiment, it is measured by solving the mass conservation of salt in the system.
In the modeling, it is calculated by estimating the diffusive and advective desalination, based on measured ice thickness $l$, bulk salinity $S_{bulk}$ and a given initial value of $\phi$.
}
\label{figs5}
\end{figure}

Fig.\ref{figs5} shows a flow chart summarizing how the average ice porosity $\phi$ is measured in the experiments and estimated with the integral desalination model. 
In the experiment, for each measurement point, we apply the mass conservation of salt in the system (Eq.\ref{eqs1}) and calculate the corresponding $\phi$.
In the model, we give an initial value of $\phi$.
For every moment, we estimate the salt flux and the rate of change for $\phi$ with the measured $l$ and $S_{bulk}$ and thus determine $\phi$ at the next moment, with Eq.\ref{eqs7} (Eq.1 in the main text):
\begin{equation}
    \begin{aligned}
        \frac{d}{dt}\left( \frac{\textcolor{blue}{S_c+S_{bulk}}}{2} \phi\right)=- \textcolor{red}{C}\cdot  \textcolor{orange}{Sh}\phi \textcolor{orange}{D}\cdot \frac{\textcolor{blue}{S_c-S_{bulk}}}{\textcolor{blue}{l^2}}.
    \end{aligned}
    \label{eqs7}
\end{equation}
More specifically, in Eq.\ref{eqs7}, the terms marked in blue are inputted by the boundary conditions or extracted values from the experiments, the terms marked in orange are inputted by the empirical formulation/values obtained from the previous work, and the term marked in red is the coefficient $C$ fitted from the experimental data, which is introduced to account for the non-even shape of the ice front and the non-uniform distribution of local porosity.
$S_c$ is the salinity at the cold plate, determined by the cold plate temperature through $T_0(S_c)=T_c=-12.1 ^\circ$C where $T_0$ represents the freezing point.
$S_{bulk}$ is the bulk salty water salinity and $l$ is the average ice thickness, both extracted from experiments.
$D$ is the salt diffusivity.
$Sh$ is the Sherwood number representing the dimensionless mass transfer efficiency due to convection in mushy ice.
The parameter $C$ is obtained by fitting the model results with the experiment results, which compensates for influences of non-uniform ice thickness and porosity, as discussed previously.
Then, with a given initial value of porosity $\phi$, the temporal evolution of $\phi$ can be explicitly solved by the model, with the Sherwood number $Sh$ estimated using the approach described in End Matter.

By comparing the $\phi$ as a function of time from experiment and model, we show that the observed aging of ice (decrease in $\phi$) in the experiment is driven primarily by diffusive and also minorly by convective desalination of ice as described by the integral desalination model.

\end{document}